\documentclass[a4paper]{article}
\usepackage{graphicx}
\usepackage[compress]{cite}
\usepackage{braket}
\usepackage{bm}
\usepackage{comment}
\usepackage{here}
\usepackage{a4wide}
\usepackage{authblk}
\begin{document}
\def\Journal#1#2#3#4{{#1} {\bf #2}, #3 (#4)}
\def\AHEP{Advances in High Energy Physics.} 	
\def\ARNPS{Annu. Rev. Nucl. Part. Sci.} 
\def\AandA{Astron. Astrophys.} 
\def\ANP{Ann. Phys.}
\def\APJ{Astrophys. J.}
\def\APJS{Astrophys. J. Suppl}
\def\COMR{Comptes Rendues}
\def\CQG{Class. Quantum Grav.}
\def\CPC{Chin. Phys. C}
\def\EPJC{Eur. Phys. J. C}
\def\EPL{EPL}
\def\FP{Fortsch. Phys.}
\def\IJMPA{Int. J. Mod. Phys. A}
\def\IJMPE{Int. J. Mod. Phys. E}
\def\JCAP{J. Cosmol. Astropart. Phys.}
\def\JHEP{J. High Energy Phys.}
\def\JETPL{JETP. Lett.}
\def\JETPUSSR{JETP (USSR)}
\def\JPG{J. Phys. G} 
\def\JPCS{J. Phys. Conf. Ser.} 
\def\JPGNP{J. Phys. G: Nucl. Part. Phys.} 
\def\JMP{J. Mod. Phys.} 
\def\MPLA{Mod. Phys. Lett. A}
\def\NIMA{Nucl. Instrum. Meth. A.}
\def\NATU{Nature}
\def\NCA{Nuovo Cimento}
\def\NJP{New. J. Phys.}
\def\NPB{Nucl. Phys. B}
\def\NPBOLD{Nucl. Phys.}
\def\NPBSUPPL{Nucl. Phys. B. Proc. Suppl.}
\def\PL{Phys. Lett.}
\def\PLB{{Phys. Lett.} B}
\def\PMCA{PMC Phys. A}
\def\PREP{Phys. Rep.}
\def\PPNP{Prog. Part. Nucl. Phys.}
\def\PLBOLD{Phys. Lett.}
\def\PAN{Phys. Atom. Nucl.}
\def\PRL{Phys. Rev. Lett.}
\def\PRD{Phys. Rev. D}
\def\PRC{Phys. Rev. C}
\def\PR{Phys. Rev.}
\def\PTP{Prog. Theor. Phys.}
\def\PTEP{Prog. Theor. Exp. Phys.}
\def\RMP{Rev. Mod. Phys.}
\def\RPP{Rep. Prog. Phys.}
\def\SJNP{Sov. J. Nucl. Phys.}
\def\SPJETP{Sov. Phys. JETP.}
\def\SCIENCE{Science}
\def\TNYAS{Trans. New York Acad. Sci.}
\def\ZETP{Zh. Eksp. Teor. Piz.}
\def\ZFPH{Z. fur Physik}
\def\ZPC{Z. Phys. C}
\title{Neutrino mixing model for best-fit values of $\theta_{12}$ and $\theta_{13}$}
\author[1,2]{Yuta Hyodo \footnote{Corresponding author: 2ctad004@mail.u-tokai.ac.jp}}
\author[3]{Teruyuki Kitabayashi\footnote{teruyuki@tokai-u.jp}}
\affil[1]{Graduate School of Science and Technology, Tokai University, 4-1-1 Kitakaname, Hiratsuka, Kanagawa 259-1292, Japan}
\affil[2]{Micro/Nano Technology Center, Tokai University, 4-1-1 Kitakaname, Hiratsuka, Kanagawa 259-1292, Japan}
\affil[3]{Department of Physics, Tokai University, 4-1-1 Kitakaname, Hiratsuka, Kanagawa 259-1292, Japan}
\date{}
\maketitle
\begin{abstract}
Recently, the precise observations have yielded values for the neutrino mixing angles, denoted as $\theta_{12}$ and $\theta_{13}$. Therefore, constructing a neutrino mixing model capable of accurately reflecting a these measurements is crucial. In this paper, we propose a novel neutrino mixing model similar to the trimaximal mixing model. Unlike the trimaximal model that falls short of concurrently replicating the best-fit values of $\theta_{12}$ and $\theta_{13}$, our proposed mixing model has the ability simultaneously reproduce the best-fit values for both angles.
\end{abstract}
\section{Introduction\label{section:introduction}}
Owing to the remarkable advancements in recent experimental technology, the precision of neutrino mixing angles measurements, specifically the solar mixing angle ($\theta_{12}$) and the reactor mixing angle ($\theta_{13}$), has substantially improves. Consequently, it is crucial to construct a neutrino mixing model capable of accurately predicting these observed values.

Various neutrino mixing models have been proposed \cite{King2008PLB,Roi2013JMP,Bazzocchi2013FP,Garg2013JHEP,Kitabayashi2013arXiv} that can simultaneously predict the best-fit values for solar and reactor mixing angles.  Conversely, models such as, the tri-bimaximal-reactor (TBR) mixing \cite{King2009PLB}, first trimaximal  (TM1) mixing \cite{Xing2007PLB,Albright2009EPJC,Albright2010EPJC}, second trimaximal (TM2) mixing\cite{Albright2009EPJC,Harrison2006PRD,Zee2007PLB,Albright2010EPJC}, and the model in Ref.\cite{Zhou2012NPB} are capable of predicting within the $3\sigma$ region for these angles; however, they fail to simultaneously predict the best-fit values.

Notably, a neutrino mixing model that does not predict the best-fit values for both the angle does not lack utility. For instance, the tri-bimaximal mixing (TBM) model fails to align with the current experimental values, yet its matrix form is very simple, and its underlying symmetries and physics are well-understood \cite{Harrison2002PLB,Xing2002PLB,LamPRD2006,LamPLB2007,LamPRL2008,Scott2002PLB,Kitabayashi2007PRD,Zee2007PLB,Albright2009EPJC,Albright2010EPJC,Altarelli2010RMP,Xing2007PLB,Harrison2006PRD}. If a slightly modified TBM model can simultaneously predict the best-fit values for both the angles, it may be considered a promising neutrino mixing model. The modified TBM model has been explored in Ref.\cite{King2008PLB,Roi2013JMP,Bazzocchi2013FP,Garg2013JHEP,King2009PLB,Xing2007PLB,Albright2009EPJC,Albright2010EPJC,Harrison2006PRD,Zee2007PLB,lashin2012PRD,Petcov2012PLB,Wu2012PLB,Siyeon2012EPJC,Duarah2012arXiv,Singh2014JPCS,Ahn2022PRD,Chao2013JHEP,Garg2018NPB,Araki2014PTEP,Dev2012PLB,Zhou2012arXiv,Jora2013IJMPA,Damanik2013CJP,Zhao2017IJMPA,Takasugi2015PTEP,Kitabayashi2013arXiv}.

This study introduces a new neutrino mixing model achieved by modifying the TBM to simultaneously predict the best-fit values of solar and reactor mixing angles. Owing to TBM's inability to simultaneously predict the two angles, the modification is conducted in two steps.  Initially, we predict the best-fit value for the solar mixing angle, followed by adjustments to enhance the prediction of the reactor mixing angle. Consequently, a new neutrino mixing model akin to the well-known trimaximal (TM) model are obtained. While the TM model fails to precisely replicate the best-fit values of the solar and reactor mixing angles, our TM-like neutrino mixing model can simultaneously attain optimal values for both angles.

The subsequent sections are organized as follows: in Sect. \ref{section:TM}, we detail the modification of the TBM to create a new mixing model capable of simultaneously reproducing the best-fit values of the solar and reactor mixing angles. Sect. \ref{section:comparison} presents a comparison of our neutrino model, derived from a modified TM model, with the original TM model. In Sect. \ref{section:Z2}, we investigate whether the neutrino mass matrix from our neutrino mixing model retains invariance under $Z_2$ symmetry, analogous to the neutrino mass matrix from TM mixing model. Sect. \ref{section:summary} provides a summary of our findings.

\section{TM-like mixing\label{section:TM}}
\subsection{Tri-bimaximal (TBM) mixing}
The neutrino mixing state is commonly expressed using the Pontecorvo-Maki-Nakagawa-Sakata mixing matrix \cite{Pontecorvo1957,Pontecorvo1958,Maki1962PTP,PDG}, as shown in
\begin{eqnarray}
U  
=\left ( 
\begin{array}{ccc}
U_{e1}&U_{e2}&U_{e3}\\
U_{\mu1}&U_{\mu2}&U_{\mu3}\\
U_{\tau1}&U_{\tau2}&U_{\tau3}
\end{array}
\right)
=
\left ( 
\begin{array}{ccc}
c_{12}c_{13} & s_{12}c_{13} & s_{13} e^{-i\delta} \\
- s_{12}c_{23} - c_{12}s_{23}s_{13} e^{i\delta} & c_{12}c_{23} - s_{12}s_{23}s_{13}e^{i\delta} & s_{23}c_{13} \\
s_{12}s_{23} - c_{12}c_{23}s_{13}e^{i\delta} & - c_{12}s_{23} - s_{12}c_{23}s_{13}e^{i\delta} & c_{23}c_{13} \\
\end{array}
\right),
\end{eqnarray}
where $c_{ij}=\cos\theta_{ij}$, $s_{ij}=\sin\theta_{ij}$ ($i,j$=1,2,3), $\theta_{ij}$ denotes the mixing angles, and $\delta$ denotes the Dirac CP phase. The sines and cosines of the three mixing angles of the PMNS matrix $U$ are expressed as follows:
\begin{eqnarray}
s_{12}^2&=&\frac{|U_{e2}|^2}{1-|U_{e3}|^2},~~ s_{23}^2=\frac{|U_{\mu3}|^2}{1-|U_{e3}|^2},~~s_{13}^2=|U_{e3}|^2,\nonumber \\
c_{12}^2&=&\frac{|U_{e1}|^2}{1-|U_{e3}|^2},~~c_{23}^2=\frac{|U_{\tau3}|^2}{1-|U_{e3}|^2}.
\label{Eq:mixing_angle_PMNS}
\end{eqnarray}
The Jarlskog rephasing invariant\cite{Jarlskog},
\begin{eqnarray}
J=Im(U_{e1}U_{e2}^{\ast}U_{\mu1}^{\ast}U_{\mu2})=s_{12}s_{23}s_{13}c_{12}c_{23}c_{13}^2\sin{\delta}
\label{Eq:Jarlskog}
\end{eqnarray}
is useful in calculating the Dirac CP phase $\delta$.

A global analysis of current data from neutrino oscillation experiments reveals the following best-fit values of the mixing angles in the case of normal mass ordering (NO), $m_1 < m_2<m_3$, where $m_1,m_2,m_3$ are the neutrino mass eigenstates,\cite{NuFit}:
\begin{eqnarray} 
s_{12}^2&=& 0.303^{+0.012}_{-0.012} \quad (0.270 \sim 0.341), \nonumber \\
s_{23}^2&=& 0.451^{+0.019}_{-0.016} \quad (0.408 \sim 0.603), \nonumber \\
s_{13}^2&=& 0.02225^{+0.00056}_{-0.00059} \quad (0.02052 \sim 0.02398), \nonumber \\
\delta/^\circ &=& 232^{+36}_{-26} \quad (144 \sim 350), 
\label{Eq:neutrino_observation_NO}
\end{eqnarray}
where $\pm$ denotes the $1 \sigma$ region, and the parentheses denote the $3 \sigma$ region. For inverted mass ordering (IO), $m_3 < m_1<m_2$, a global analysis reveals the following \cite{NuFit}:
\begin{eqnarray} 
s_{12}^2&=& 0.303^{+0.012}_{-0.012} \quad (0.270 \sim 0.341), \nonumber \\
s_{23}^2&=& 0.569^{+0.016}_{-0.021} \quad (0.412 \sim 0.613), \nonumber \\
s_{13}^2&=& 0.02223^{+0.00058}_{-0.00058} \quad (0.02048 \sim 0.02416), \nonumber \\
\delta/^\circ &=& 276^{+22}_{-29} \quad (194  \sim 344).
\label{Eq:neutrino_observation_IO}
\end{eqnarray}
The best-fit values and $3 \sigma$ regions of $s_{13}^2$ are nearly identical for both the case of NO and IO. Consequently, we consider the observed best-fit values and $3 \sigma$ region of $s_{13}^2$ for the NO case.

The mixing matrix under investigation in this study represented by the TBM matrix, denoted as follows, as originally presented in Ref.\cite{Harrison2002PLB,Xing2002PLB}:
\begin{eqnarray}
U_{\rm TBM}=\left(
\begin{array}{ccc}
\sqrt{\frac{2}{3}} & \sqrt{\frac{1}{3}}  & 0 \\
-\sqrt{\frac{1}{6}} &  \sqrt{\frac{1}{3}}  &\sqrt{\frac{1}{2}}  \\
\sqrt{\frac{1}{6}} &  -\sqrt{\frac{1}{3}}  &\sqrt{\frac{1}{2}}
\end{array}
\right).
\label{Eq:UTBM}
\end{eqnarray}
The values of $s_{12}^2$ and $s_{13}^2$ obtained from the TBM, as indicated in
\begin{eqnarray}
(s_{12}^2)_{\rm TBM}= |U_{e2}|^2 = \frac{1}{3}=0.3333, \quad  
(s_{13}^2)_{\rm TBM}=|U_{e3}|^2 = 0,
\label{Eq:TBM_Predicted}
\end{eqnarray}
do not match the best-fit values of $s_{12}^2 =0.303$ and $s_{13}^2 = 0.02225$.

\subsection{Improved reproducibility of the solar mixing angle}
First, the TBM matrix is adjusted to achieve the reproduction of the best-fit values for $s_{12}^2$. Referring to Eq.(\ref{Eq:TBM_Predicted}), if $|U_{e2}|^2$ is made equal to the best-fit values of $s_{12}^2$, an enhancement in reproducibility is achieved. We introduce the modification to $U_{e2}$ as follows:
\begin{eqnarray}
\sqrt{\frac{1}{3}} \rightarrow \sqrt{\frac{1}{3}+\alpha},
\end{eqnarray}
where $\alpha$ denotes the real parameter. When $\alpha=-0.0303$, the best-fit value $s_{12}^2 = 0.303$ is reproduced. The modified neutrino mixing matrix is represented as follows: 
\begin{eqnarray}
U_{\rm MTBM}^{0}=\left(
\begin{array}{ccc}
\sqrt{\frac{2}{3}} & \sqrt{\frac{1}{3} + \alpha}  & 0 \\
-\sqrt{\frac{1}{6}} &  \sqrt{\frac{1}{3}}  &\sqrt{\frac{1}{2}}  \\
\sqrt{\frac{1}{6}} &  -\sqrt{\frac{1}{3}}  &\sqrt{\frac{1}{2}}
\end{array}
\right).
\label{Eq:UMTBM_non}
\end{eqnarray}

The best-fit value of $s_{12}^2$ is derived from $U_{\rm MTBM}^{0}$; however, $U_{\rm MTBM}^{0}$ is not a unitary matrix. To satisfy the unitarity condition $U_{\rm MTBM}^\dag U_{\rm MTBM}=1$, we modified Eq. (\ref{Eq:UMTBM_non}) to
\begin{eqnarray}
U_{\rm MTBM}=\left(
\begin{array}{ccc}
\sqrt{\frac{2}{3} + x_1} & \sqrt{\frac{1}{3} + \alpha}  & 0 \\
-\sqrt{\frac{1}{6}+ x_2} &  \sqrt{\frac{1}{3}+ x_4}  &\sqrt{\frac{1}{2}+ x_6}  \\
\sqrt{\frac{1}{6}+ x_3} &  -\sqrt{\frac{1}{3}+ x_5}  &\sqrt{\frac{1}{2}+ x_7}
\end{array}
\right)
\label{Eq:UMTBM_unitarity}
\end{eqnarray}
and subsequently determined the values of the real parameters $x_i$ ($i=1,2,\cdots,7$). It is equivalent to finding the solution for the system of simultaneous equations denoted by
\begin{eqnarray}
&& 1+x_1+\alpha = 1, \quad 1+x_2+x_4+x_6=1,  \quad 1+x_3+x_5+x_7=1, \nonumber \\
&& -\sqrt{\left(\frac{2}{3}+x_1\right) \left(\frac{1}{6}+x_2\right)}+\sqrt{\left(\frac{1}{3}+\alpha\right) \left(\frac{1}{3}+x_4\right)}=0,  \nonumber \\
&& \sqrt{\left(\frac{2}{3}+x_1\right)\left(\frac{1}{6}+x_3\right)}-\sqrt{\left(\frac{1}{3}+\alpha\right)\left(\frac{1}{3}+x_5\right)}=0,\nonumber \\
&& -\sqrt{\left(\frac{1}{6}+x_2\right)\left(\frac{1}{6}+x_3\right)}-\sqrt{\left(\frac{1}{3}+x_4\right)\left(\frac{1}{3}+x_5\right)}+\sqrt{\left(\frac{1}{2}+x_6\right)\left(\frac{1}{2}+x_7\right)}=0. 
\end{eqnarray}
Since there are six independent equations but seven variables, the simultaneous equations cannot be solved in their current form. We focus on 
\begin{eqnarray}
(s_{23}^2)_{\rm TBM}=|U_{\mu3}|^2=\frac{1}{2},
\end{eqnarray}
which is derived from Eq.(\ref{Eq:mixing_angle_PMNS}) in the TBM and is consistent with the observation.
Therefore, we can derive
\begin{eqnarray}
x_6=0.
\end{eqnarray}
The solution to the simultaneous equations in this case is denoted as follows:
\begin{eqnarray}
(x_1,x_2,x_3,x_4,x_5,x_7) =\left (-\alpha,\frac{\alpha}{2},\frac{\alpha}{2},-\frac{\alpha}{2},-\frac{\alpha}{2},0\right).
\label{Eq:solution}
\end{eqnarray}
By adopting the solution presented in Eq.(\ref{Eq:solution}), Eq.(\ref{Eq:UMTBM_non}) becomes
\begin{eqnarray}
U_{\rm MTBM}=\left(
\begin{array}{ccc}
\sqrt{\frac{2}{3} - \alpha} & \sqrt{\frac{1}{3} + \alpha}  & 0 \\
-\sqrt{\frac{1}{6}+ \frac{\alpha}{2}} &  \sqrt{\frac{1}{3}-\frac{\alpha}{2}}  &\sqrt{\frac{1}{2}}  \\
\sqrt{\frac{1}{6}+ \frac{\alpha}{2}} &  -\sqrt{\frac{1}{3}-\frac{\alpha}{2}}  &\sqrt{\frac{1}{2}}
\end{array}
\right).
\label{Eq:UMTBM}
\end{eqnarray}

The range of $\alpha$ is determined by
\begin{eqnarray}
-\frac{1}{3} < \alpha < \frac{2}{3}
\label{Eq:a_region_1}
\end{eqnarray}
because the contents of the square root cannot be negative.

\subsection{Improved reproducibility of the reactor mixing angle}
Subsequently, we modified TBM to enable the reproduction of the best-fit values for the reactor mixing angle. In this study, we adopt the method used in Ref. \cite{Xing2007PLB} to modify the TBM and obtain a non-zero reactor mixing angle.

In Ref. \cite{Xing2007PLB}, the TM1 mixing matrix is derived, wherein the first column of the TBM mixing matrix remains unchanged, while the second and third columns are adjusted. Similarly, we consider a mixing matrix that preserves the first column of $U_{\rm MTBM}$ as follows:

\begin{eqnarray}
U_{\rm MTM1}&=&U_{\rm MTBM}U_{23}\nonumber\\
&=&\left(
\begin{array}{ccc}
\sqrt{\frac{2-3\alpha}{3}} &\sqrt{\frac{1}{3} + \alpha}\cos{\theta}&\sqrt{\frac{1}{3} + \alpha}\sin{\theta}e^{-i\phi} \\
-\sqrt{\frac{1}{6}+ \frac{\alpha}{2}}&  \sqrt{\frac{1}{3}-\frac{\alpha}{2}}\cos{\theta}-\frac{\sin{\theta}}{\sqrt{2}}e^{i\phi}& \frac{\cos{\theta}}{\sqrt{2}}+ \sqrt{\frac{1}{3}-\frac{\alpha}{2}}\sin{\theta}e^{-i\phi} \\
\sqrt{\frac{1}{6}+ \frac{\alpha}{2}}& - \sqrt{\frac{1}{3}-\frac{\alpha}{2}}\cos{\theta} -\frac{\sin{\theta}}{\sqrt{2}}e^{i\phi}&\frac{\cos{\theta}}{\sqrt{2}}-\sqrt{\frac{1}{3}-\frac{\alpha}{2}}\sin{\theta}e^{-i\phi}
\end{array}
\right),
\label{Eq:UMTM1}
\end{eqnarray}
where
\begin{eqnarray}
U_{23}=\left(
\begin{array}{ccc}
1&0&0\\
0&\cos{\theta}&\sin{\theta}e^{-i\phi}\\
0&-\sin{\theta}e^{i\phi}&\cos{\theta}
\end{array}
\right),
\label{Eq:U23}
\end{eqnarray}
$\theta$ denotes a rotation angle, and $\phi$ denotes a phase parameter. We refer to Eq. (\ref{Eq:UMTM1}) as MTM1. From Eq. (\ref{Eq:mixing_angle_PMNS}), the mixing angles of MTM1 mixing are expressed as follows:
\begin{eqnarray}
s_{13}^2=\left(\frac{1}{3}+\alpha\right)\sin^2{\theta},~~s_{12}^2=\frac{\left(1+3\alpha \right)\cos^2{\theta}}{3-\left(1+3\alpha \right)\sin^2{\theta}},\nonumber\\
s_{23}^2=\frac{3\cos^2{\theta}+\left(2-3\alpha \right)\sin^2{\theta}+\sqrt{3\left(2-3\alpha \right)}\cos{\phi}\sin{2\theta}}{6-2\left(1+3\alpha \right)\sin^2{\theta}},
\label{Eq:mixing_angle_NTM1}
\end{eqnarray}
with $\theta$ and $\phi$. 

A benchmark point
\begin{eqnarray}
(\alpha, \theta, \phi) = (-0.01483,0.2675~\rm{rad}, 1.571~ \rm{rad})
\label{Eq:NTM1_p}
\end{eqnarray}
yields
\begin{eqnarray}
(s_{12}^2,s_{13}^2,s_{23}^2) = (0.3030, 0.02225, 0.5000).
\label{Eq:NTM1_mixing_angle}
\end{eqnarray}
The best-fit values for the solar and reactor mixing angles can be simultaneously reproduced.

In Ref.\cite{Harrison2006PRD}, the TM2 mixing matrix is established, wherein the second column of the TBM mixing matrix remains unaltered, while the first and third columns are modified. Similarly, we contemplate a mixing matrix that retains the second column of $U_{\rm MTBM}$ unchanged as follows:
\begin{eqnarray}
U_{\rm MTM2}&=&U_{\rm MTBM}U_{13}\nonumber\\
&=&\left(
\begin{array}{ccc}
\sqrt{\frac{2}{3}-a}\cos{\theta}&\sqrt{\frac{1}{3}+a}&\sqrt{\frac{2}{3}-\alpha}\sin{\theta}e^{-i\phi}\\
-\sqrt{\frac{1}{6}+\frac{\alpha}{2}}\cos{\theta}-\frac{\sin{\theta}e^{i\phi}}{\sqrt{2}}&\sqrt{\frac{1}{3}-\frac{a}{2}}&\frac{\cos{\theta}}{\sqrt{2}}-\sqrt{\frac{1}{6}+\frac{\alpha}{2}}\sin{\theta}e^{-i\phi}\\
\sqrt{\frac{1}{6}+\frac{\alpha}{2}}\cos{\theta}-\frac{\sin{\theta}e^{i\phi}}{\sqrt{2}}&-\sqrt{\frac{1}{3}-\frac{a}{2}}&\frac{\cos{\theta}}{\sqrt{2}}+\sqrt{\frac{1}{6}+\frac{\alpha}{2}}\sin{\theta}e^{-i\phi}
\end{array}
\right)
\label{Eq:UMTM2}
\end{eqnarray}
where
\begin{eqnarray}
U_{13}=\left(
\begin{array}{ccc}
\cos{\theta}&0&\sin{\theta}e^{-i\phi}\\
0&1&0\\
-\sin{\theta}e^{i\phi}&0&\cos{\theta}
\end{array}
\right).
\label{Eq:U13}
\end{eqnarray}
We refer to Eq. (\ref{Eq:UMTM2}) as MTM2. The mixing angles of MTM2 mixing are expressed as follows:
\begin{eqnarray}
s_{13}^2=\left(\frac{2}{3}-\alpha \right)\sin^2{\theta},~~s_{12}^2=\frac{1+3\alpha}{3-\left(2-3\alpha\right)\sin^2{\theta}},\nonumber\\
s_{23}^2=\frac{3\cos^2{\theta}+\left(1+3\alpha \right)\sin^2{\theta}-\sqrt{3\left(1+3\alpha \right)}\sin{2\theta}\cos{\phi}}{6-2(2-3\alpha)\sin^2{\theta}}.
\label{Eq:mixing_angle_NTM2}
\end{eqnarray}
A benchmark point
\begin{eqnarray}
(\alpha, \theta, \phi) =  (-0.03708,0.1788~ \rm{rad}, 1.571~ \rm{rad})
\label{Eq:NTM2_p}
\end{eqnarray}
yields
\begin{eqnarray}
(s_{12}^2,s_{13}^2,s_{23}^2) = (0.3030, 0.02226, 0.5000).
\label{Eq:NTM2_mixing_angle}
\end{eqnarray}
The best-fit values for both angles can be simultaneously reproduced.
\section{Comparison of TM and MTM\label{section:comparison}}
\begin{figure}[t]
\begin{tabular}{cc}
\hspace{-3.8cm}
\begin{minipage}[t]{0.3\hsize}
\centering
\includegraphics[keepaspectratio, scale=1.0]{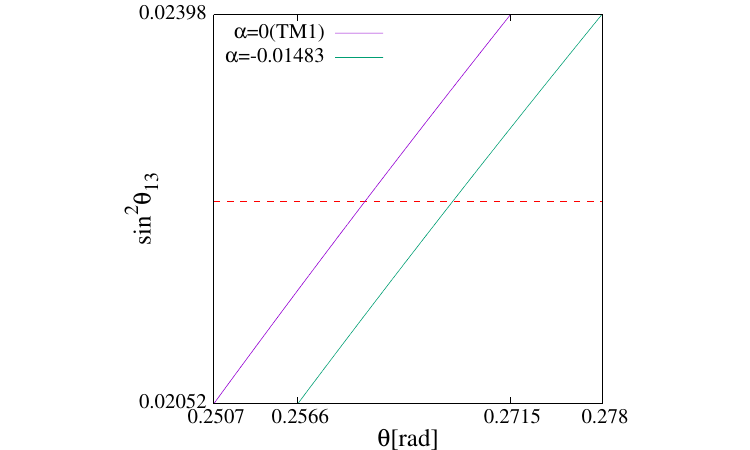}
\end{minipage} &\hspace{0.3\columnwidth}
\begin{minipage}[t]{0.3\hsize}
\centering
\includegraphics[keepaspectratio, scale=1.0]{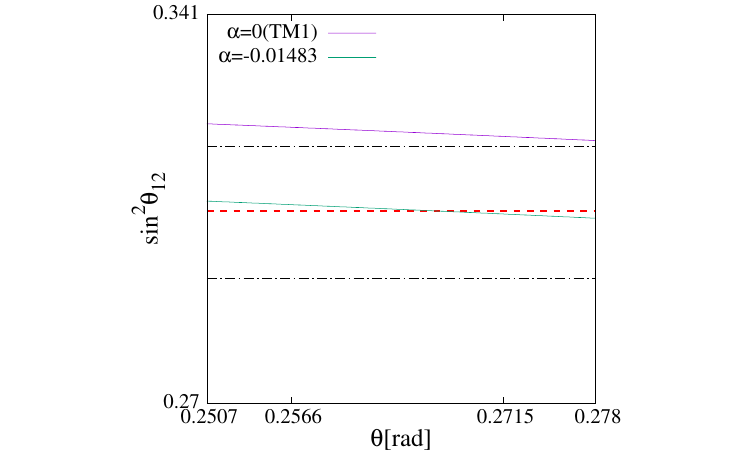}
 \end{minipage} \\
\end{tabular}
 \caption{Dependence of the mixing angles $s_{13}^2$ and $s_{12}^2$ on rotation angle $\theta$ in rad for TM1 ($\alpha=0$) and MTM1($\alpha=-0.01483$). The left (right) panel shows the relationship between $s_{13}^2$ ($s_{12}^2$) and $\theta$.}
 \label{fig:TM1_NTM1_mixing_angle}
  \end{figure}
\begin{figure}[th]
\centering
\includegraphics[keepaspectratio, scale=1.0]{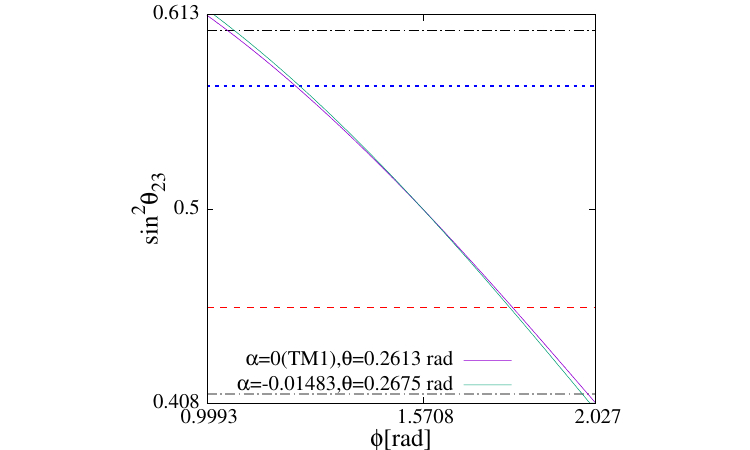}
 \caption{Dependence of the mixing angle $s_{23}^2$ on phase parameter $\phi$ in rad for TM1 ($\alpha=0$) and MTM1($\alpha=-0.01483$). We choose $\theta$, such that $s_{13}^2$ becomes the best-fit value.}
 \label{fig:TM1_NTM1_mixing_angle2}
  \end{figure}
\begin{figure}[th]
\begin{tabular}{cc}
\hspace{-3.8cm}
\begin{minipage}[t]{0.3\hsize}
\centering
\includegraphics[keepaspectratio, scale=1.0]{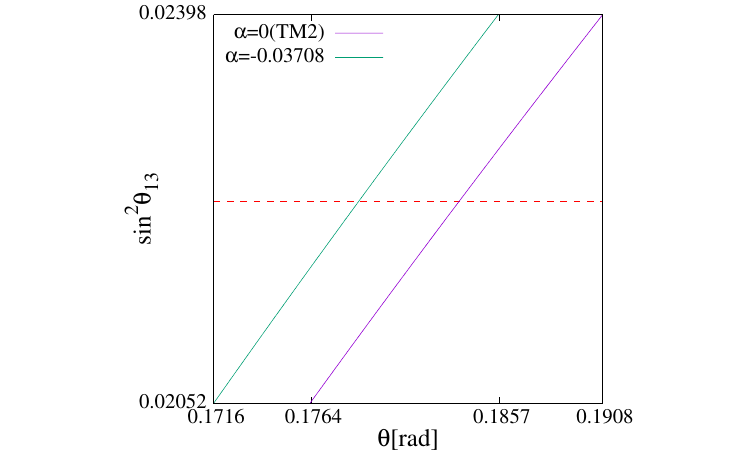}
\end{minipage} &\hspace{0.3\columnwidth}
\begin{minipage}[t]{0.3\hsize}
\centering
\includegraphics[keepaspectratio, scale=1.0]{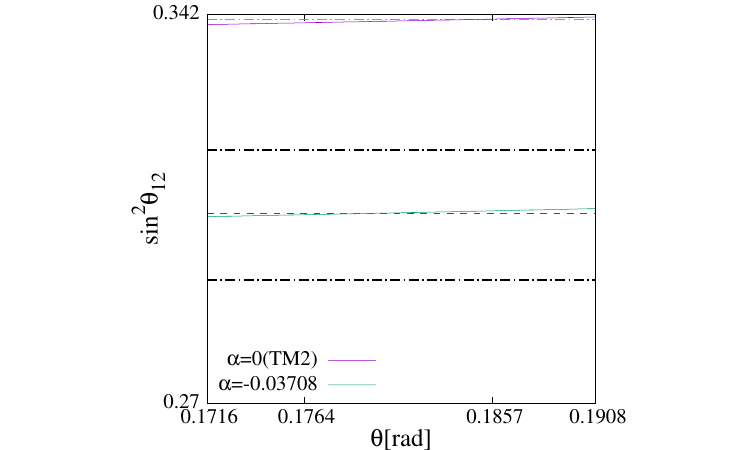}
 \end{minipage} \\
\end{tabular}
 \caption{Dependence of the mixing angles $s_{13}^2$ and $s_{12}^2$ on rotation angle $\theta$ in rad for TM2 ($\alpha=0$) and MTM2 ($\alpha=-0.03708$). The left (right) panel shows the relationship between $s_{13}^2$ ($s_{12}^2$) and $\theta$.}
 \label{fig:TM2_NTM2_mixing_angle}
  \end{figure}
\begin{figure}[th]
\centering
\includegraphics[keepaspectratio, scale=1.0]{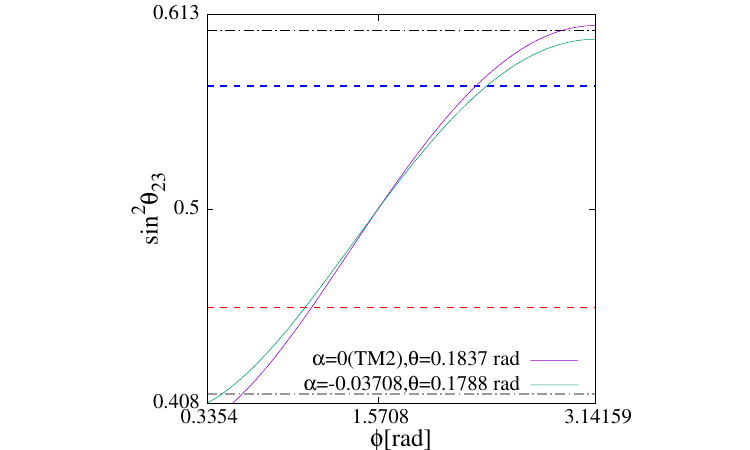}
 \caption{Dependence of the mixing angle $s_{23}^2$ on phase parameter $\phi$ in rad for TM2 ($\alpha=0$) and MTM2 ($\alpha=-0.03708$). We choose $\theta$, such that $s_{13}^2$ becomes the best-fit value.}
 \label{fig:TM2_NTM2_mixing_angle2}
  \end{figure}
When $\alpha=0$, MTM1 (MTM2) is essentially equivalent to TM1 (MTM1). Therefore, MTM1 and MTM2 can be regarded as modified versions of TM1 and TM2. We will now examine the distinctions between TM1 (TM2) and MTM1 (MTM2).
\subsection{TM1 and MTM1}
Fig.\ref{fig:TM1_NTM1_mixing_angle} illustrates the dependence of the mixing angle $s_{13}^2$ and $s_{12}^2$ on the rotation angles $\theta$ in rad for TM1($\alpha=0$) and MTM1($\alpha=-0.01483$). In the left panel, the predicted values of $s_{13}^2$ obtained from TM1 ($\alpha=0$) and MTM1 ($\alpha=-0.01483$) are displayed.  The red dotted line in the figure represents the best-fit value of $s_{13}^2$. The horizontal axis indicates the range of $\theta$ where $s_{13}^2$ falls within the $3\sigma$ region. The right panel of Fig.\ref{fig:TM1_NTM1_mixing_angle} displays the dependence of $s_{12}^2$ predicted from TM1($\alpha=0$) and MTM1($\alpha=-0.01483$) on the rotation angle $\theta$. The red dotted line represents the best-fit value of $s_{12}^2$, and the black dotted line denotes the $1 \sigma$ region of $s_{12}^2$. Key observations from Fig. \ref{fig:TM1_NTM1_mixing_angle} include:
\begin{itemize}
\item For TM1 $\left(\alpha=0\right)$, when $s_{13}^2$ satisfies the best-fit value, then $s_{12}^2=0.3182$. The value of $s_{12}^2$ for TM1 closely approaches the upper limit of the $1\sigma$ region, which is $s_{12}^2=0.315$. Hence, it is not feasible to simultaneously reproduce the best-fit values of $s_{13}^2$ and $s_{12}^2$ for TM1.
\item For MTM1 $\left(\alpha=-0.01483\right)$, using the parameter in Eq.(\ref{Eq:NTM1_p}), the best-fit values of $s_{13}^2$ and $s_{12}^2$ are simultaneously reproduced.
\end{itemize}

Fig.\ref{fig:TM1_NTM1_mixing_angle2} demonstrates the dependence of $s_{23}^2$ predicted from TM1 ($\alpha=0$) and MTM1 ($\alpha=-0.01483$) on the phase parameter $\phi$. The red (blue) dotted line represents the best-fit value of $s_{23}^2$ for the NO (IO) case, and the black (gray) dotted line shows the upper (lower) boundary of the $3 \sigma$ region of $s_{23}^2$ for the NO (IO) case. We select $\theta$, such that $s_{13}^2$ matches the best-fit values. Key observations from Fig. \ref{fig:TM1_NTM1_mixing_angle2} include:
\begin{itemize}
\item In the case of TM1 $\left(\alpha=0\right)$, when $s_{13}^2$ aligns with the best-fit value, $s_{23}^2$ falls within the $3 \sigma$ region.
\item In the case of MTM1 ($\alpha=-0.01483$), if both $s_{13}^2$ and $s_{12}^2$ conform to the best-fit values, $s_{23}^2$ also falls within the $3 \sigma$ region.
\end{itemize}

Using Eq.(\ref{Eq:Jarlskog}), Jarlskog rephasing invariant of MTM1 is written as follows:
\begin{eqnarray}
J_{\rm{MTM1}}=\frac{(1+3\alpha)\sqrt{3(2-3\alpha)}}{36}\sin{2\theta}\sin{\phi}
\label{Eq:Jarlskog_NTM1}
\end{eqnarray}
We obtain the Dirac CP phase of MTM1
\begin{eqnarray}
\tan{\delta}=\frac{1+\Delta_{\rm{MTM1}}\cos{2\theta}}{\Delta_{\rm{MTM1}}+\cos{2\theta}}\tan{\phi},
\label{Eq:Delta_NTM1}
\end{eqnarray}
where $\Delta_{\rm{MTM1}}$ denotes
\begin{eqnarray}
\Delta_{\rm{MTM1}}=\frac{1+3\alpha}{5-3\alpha}.
\end{eqnarray}
For TM1 ($\alpha=0$), the values of $\Delta_{\rm{MTM1}}$ is calculated as $\Delta_{\rm{MTM1}}=\frac{1}{5}=0.2$. In contrast, for MTM1 ($\alpha=-0.01483$), the value of $\Delta_{\rm{MTM1}}$ is determined to be $\Delta_{\rm{MTM1}}=0.1894$. Consequently, the Dirac CP phase predicted by TM1 mixing, denoted as $\delta_{\rm{TM1}}$, and the Dirac CP phase predicted by MTM1, denoted as $\delta_{\rm{MTM1}}$, are nearly idntical, 
\begin{eqnarray}
\delta_{\rm{TM1}} \simeq \delta_{\rm{MTM1}}.
\end{eqnarray}
%

\subsection{TM2 and MTM2}

Fig.\ref{fig:TM2_NTM2_mixing_angle} is analogous to Fig.\ref{fig:TM1_NTM1_mixing_angle}; nonetheless, it focuses on TM2 ($\alpha=0$) and MTM2 ($\alpha=-0.03708$). The gray dotted line in the right panel represents the upper limit of the $3 \sigma$ region. Key observations from Fig.\ref{fig:TM2_NTM2_mixing_angle}:
\begin{itemize}
\item For TM2 $\left(\alpha=0\right)$, when $s_{13}^2$ conforms to the best-fit value, we observe that $s_{12}^2=0.341$. Notably, the value of $s_{12}^2$ for TM1 approaches the upper limit of the $3 \sigma$ region, $s_{12}^2=0.315$. This raises concerns regarding the potential exclusion of the TM2 mixing model in future neutrino oscillation experiments.
\item For MTM2 ($\alpha=-0.03708$), using the parameter in Eq.(\ref{Eq:NTM2_p}), the best-fit values of $s_{13}^2$ and $s_{12}^2$ are simultaneously reproduced.
\end{itemize}

Fig.\ref{fig:TM2_NTM2_mixing_angle2} parallels Fig.\ref{fig:TM1_NTM1_mixing_angle2}; nonetheless, it focusses on TM2 ($\alpha=0$) and MTM2 ($\alpha=-0.03708$). Key points to note in Fig.\ref{fig:TM2_NTM2_mixing_angle2} include:
\begin{itemize}
 \item  In the case of TM2 $\left(\alpha=0\right)$, when $s_{13}^2$ matches the best-fit value, $s_{23}^2$ falls within the $3 \sigma$ region.
 \item The upper limit of $s_{23}^2$ predicted by TM2 ($\alpha=0$) is $s_{23}^2=0.6061$.
 \item  In the case of MTM2 ($\alpha=-0.03708$), if both $s_{13}^2$ and $s_{12}^2$ conform to the best-fit values, $s_{23}^2$ also falls within the $3 \sigma$ region.
 \item The upper limit of $s_{23}^2$ predicted by MTM2 ($\alpha=-0.03708$) is $s_{23}^2=0.5974$.
\end{itemize}

Using Eq.(\ref{Eq:Jarlskog}), Jarlskog rephasing invariant of MTM2 is written as
\begin{eqnarray}
J_{\rm{MTM2}}=\frac{(2-3\alpha)\sqrt{3(1+3\alpha)}}{36}\sin{2\theta}\sin{\phi}.
\label{Eq:Jarlskog_NTM2}
\end{eqnarray}
We obtain Dirac CP phase of MTM2
\begin{eqnarray}
\tan{\delta}=\frac{1+\Delta_{\rm{MTM2}}\cos{2\theta}}{\Delta_{\rm{MTM2}}+\cos{2\theta}}\tan{\phi}.
\label{Eq:Delta_NTM2}
\end{eqnarray}
where $\Delta_{\rm{MTM2}}$ denotes
\begin{eqnarray}
\Delta_{\rm{MTM2}}=\frac{2-3\alpha}{4+3\alpha}.
\end{eqnarray}
For TM2 ($\alpha=0$), $\Delta_{\rm{MTM2}}$ becomes $\Delta_{\rm{MTM2}}=\frac{1}{2}=0.5$, whereas for MTM2 ($\alpha=-0.03708$), $\Delta_{\rm{MTM2}}$ becomes $\Delta_{\rm{MTM2}}=0.5429$. Consequently, the Dirac CP phase predicted by TM2 mixing $\delta_{\rm{TM2}}$ and the Dirac CP phase predicted by MTM2 $\delta_{\rm{MTM2}}$ are almost equal, 
\begin{eqnarray}
\delta_{\rm{TM2}} \simeq \delta_{\rm{MTM2}}.
\end{eqnarray}
%
\section{$Z_2$ symmetry\label{section:Z2}}
The discussion of symmetry is crucial in neutrino physics. Studies on discrete symmetries, including $A_4$, $S_n$, and $Z_2$, have been undertaken within the frameworks of TBM, TM1, and TM2 mixing models\cite{Altarelli2010RMP,LamPRD2006,LamPLB2007,LamPRL2008}. In this context, we now elucidate the $Z_2$ symmetry inherent in the MTM1 and MTM2 mixing models.
\subsection{Generalized hidden $Z_2$ symmetry of neutrino mixing}
Assuming that the mass matrix of charges leptons is diagonal and real, the neutrino mass matrix $M_{\nu}$ is expressed as follows:
\begin{eqnarray}
M_{\nu}=U^{\ast}\textrm{diag} (m_1,m_2,m_3)U^{\dag}.
\label{Eq:mass_matrix}
\end{eqnarray}

If the neutrino mass matrix $M_{\nu}$ satisfies the transformation
\begin{eqnarray}
G^T M_{\nu} G&=&M_{\nu},\nonumber\\
G^2&=&\textrm{diag}(1,1,1),
\label{Eq:Z2}
\end{eqnarray}
the neutrino mass matrix $M_{\nu}$ is invariant under  $Z_2$ symmetry\cite{LamPRD2006}. In this context, $G$ is written as
\begin{eqnarray}
G&=&g_1v_1v_1^{\dag}+g_2v_2v_2^{\dag}+g_3v_3v_3^{\dag},
\label{Eq:G_Z2}
\end{eqnarray}
 $v_1,v_2,v_3$ are column vectors of the mixing matrix, and one of the eigenvalues among $g_1,g_2,g_3$ is equal to $-1(1)$, while the remaining two eigenvalues are equal to $1(-1)$\cite{LamPRD2006,LamPLB2007,LamPRL2008}.

As an example, in Ref. \cite{DicusPRD2011,GePRD2011}, the mixing matrix is
\begin{eqnarray}
U_k=
\left(
\begin{array}{ccc}
\frac{-k}{\sqrt{2+k^2}}&\frac{\sqrt{2}}{\sqrt{2+k^2}}   & 0 \\
\frac{1}{\sqrt{2+k^2}}& \frac{-k}{\sqrt{2(2+k^2)}}  &-\sqrt{\frac{1}{2}}  \\
 \frac{1}{\sqrt{2+k^2}}&\frac{-k}{\sqrt{2(2+k^2)}}    &\sqrt{\frac{1}{2}}
\end{array}
\right),
\end{eqnarray}
where
\begin{eqnarray}
c_{12}=\frac{-k}{\sqrt{k^2+2}},~~s_{12}=\frac{\sqrt{2}}{\sqrt{k^2+2}}.
\end{eqnarray}
$v_i$ are the $i$-th column vectors of mixing matrix $U_k$. Assuming $g_1=1$,$g_2=g_3=-1$, and using Eq.(\ref{Eq:G_Z2}), $G$ is expressed as follows:
\begin{eqnarray}
G_k=\frac{1}{2+k^2}\left(
\begin{array}{ccc}
2-k^2&2k&2k\\
2k&k^2&-3\\
2k&-2&k^2
\end{array}
\right).
\end{eqnarray}
Since the neutrino mass matrix $M_{\nu}$
\begin{eqnarray}
M_{\nu}=
\left(
\begin{array}{ccc}
A&B_1&B_2\\
B_1&C_1&D\\
B_2&D&C_2
\end{array}
\right),
\label{Eq:NMM_Z2}
\end{eqnarray}
satisfies $G_k^T M_{\nu} G_k=M_{\nu}$, the neutrino mass matrix $M_{\nu}$ is invariant under $Z_2$ symmetry.

\subsection{$Z_2$ symmetry of MTM1 and MTM2}
The column vectors of $U_{\rm MTBM}$ are
\begin{eqnarray}
v_1&=&\left(\sqrt{\frac{2}{3}-\alpha},-\sqrt{\frac{1}{6}+\frac{\alpha}{2}},\sqrt{\frac{1}{6}+\frac{\alpha}{2}}\right)^T\nonumber\\
v_2&=&\left(\sqrt{\frac{1}{3} + \alpha} , \sqrt{\frac{1}{3}-\frac{\alpha}{2}},-\sqrt{\frac{1}{3}-\frac{\alpha}{2}}\right)^T\nonumber\\
v_3&=&\left(0,\sqrt{\frac{1}{2}},\sqrt{\frac{1}{2}}\right)^T.
\label{Eq:column_vector_MTBM}
\end{eqnarray}
Based on Eq.(\ref{Eq:G_Z2}), $G_i~~(i=1,2,3)$ can be expressed as
\begin{eqnarray}
G_1&=&-v_1\left(v_1\right)^{\dag}+v_2\left(v_2\right)^{\dag}+v_3\left(v_3\right)^{\dag}\nonumber \\
&=&\frac{1}{3}\left(
\begin{array}{ccc}
-1+6\alpha&\sqrt{4+6\alpha(1-3\alpha)}&-\sqrt{4+6\alpha(1-3\alpha)}\\
\sqrt{4+6\alpha(1-3\alpha)}&2-3\alpha&1+3\alpha\\
-\sqrt{4+6\alpha(1-3\alpha)}&1+3\alpha&2-3\alpha
\end{array}
\right), 
\end{eqnarray}
\begin{eqnarray}
G_2&=&v_1\left(v_1\right)^{\dag}-v_2\left(v_2\right)^{\dag}+v_3\left(v_3\right)^{\dag}\nonumber \\
&=&\frac{1}{3}\left(
\begin{array}{ccc}
1-6\alpha&-\sqrt{4+6\alpha(1-3\alpha)}&\sqrt{4+6\alpha(1-3\alpha)}\\
-\sqrt{4+6\alpha(1-3\alpha)}&1+3\alpha&2-3\alpha\\
\sqrt{4+6\alpha(1-3\alpha)}&2-3\alpha&1+3\alpha
\end{array}
\right),
\end{eqnarray}
and,
\begin{eqnarray}
G_3=v_1\left(v_1\right)^{\dag}+v_2\left(v_2\right)^{\dag}-v_3\left(v_3\right)^{\dag}
=\left(
\begin{array}{ccc}
1&0&0\\
0&0&-1\\
0&-1&0
\end{array}
\right),
\end{eqnarray}
where $G_1^2=G_2^2=G_3^2=\textrm{diag}(1,1,1)$.

The neutrino mass matrix $M_{\rm{MTM1}}$ obtained from $U_{\rm MTM1}$ is 
\begin{eqnarray}
M_{\rm{MTM1}}&=&U^{\ast}_{\rm{MTM1}}\textrm{diag}\left(m_1,m_2,m_3\right)U^{\dag}_{\rm{MTM1}}\nonumber\\
&=&\left(
\begin{array}{ccc}
a&b&c\\
b&d&-a+d+\frac{-c(1+3\alpha)-3b(1-3\alpha)}{\sqrt{4+6\alpha(1-3\alpha)}}\\
c&-a+d+\frac{-c(1+3\alpha)-3b(1-3\alpha)}{\sqrt{4+6\alpha(1-3\alpha)}}&d-\frac{(b+c)\sqrt{2(2-3\alpha)}}{\sqrt{1+3\alpha}}
\end{array}
\right).
\label{Eq:M_NTM1}
\end{eqnarray}
The neutrino mass matrix $M_{\rm{MTM2}}$ obtained from $U_{\rm MTM2}$ is 
\begin{eqnarray}
M_{\rm{MTM2}}&=&U^{\ast}_{\rm{MTM2}}\textrm{diag}(m_1,m_2,m_3)U^{\dag}_{\rm{MTM2}}\nonumber\\
&=&\left(
\begin{array}{ccc}
a&b&c\\
b&d&-a+d+\frac{2c+3\alpha(3b-c)}{\sqrt{4+6\alpha(1-3\alpha)}}\\
c&-a+d+\frac{2c+3\alpha(3b-c)}{\sqrt{4+6\alpha(1-3\alpha)}}&d+\frac{(b+c)\sqrt{4+6\alpha(1+3\alpha)}}{2-3\alpha}
\end{array}
\right).
\label{Eq:M_NTM2}
\end{eqnarray}
Since these neutrino mass matrices are satisfied through transformation 
\begin{eqnarray}
G_1^T M_{\rm{MTM1}}G_1=M_{\rm{MTM1}}
\label{Eq:tansformation_NTM1}
\end{eqnarray}
and
\begin{eqnarray}
G_2^T M_{\rm{MTM2}}G_2=M_{\rm{MTM2}},
\label{Eq:tansformation_NTM2}
\end{eqnarray}
they are invariant under $Z_2$ symmetry.

\section{Summary\label{section:summary}}
Owing to the remarkable advancements in recent experimental technology, the measurement of neutrino mixing has reached a high level of precision. Consequently, it becomes imperative to construct a neutrino mixing model capable of accurately predicting the observed values of the solar and reactor mixing angles.

In this study, we developed a novel neutrino mixing model that can simultaneously predict the best-fit values of the solar and reactor mixing angles with a slight modification of the TBM. Notably, the values of both the angles obtained from the TBM do not match with the best-fit values. To address this issue, a modification of the TBM is performed in two steps. 

First, to reproduce the best-fit value of the solar mixing angle, we constructed $U_{\rm{MTBM}}$ by modifying
\begin{eqnarray}
\sqrt{\frac{1}{3}} \rightarrow \sqrt{\frac{1}{3}+\alpha},
\end{eqnarray}
where $\alpha$ denotes the real parameter.

Next, to improve the reproducibility of the reactor mixing angle, we modified $U_{\rm{MTBM}}$ to $U_{\rm{MTM1}}=U_{\rm{MTBM}}U_{23}$ ($U_{\rm{MTM2}}=U_{\rm{MTBM}}U_{13}$), where $U_{23}$ and $U_{13}$ denote 
\begin{eqnarray}
U_{23}=\left(
\begin{array}{ccc}
1&0&0\\
0&\cos{\theta}&\sin{\theta}e^{-i\phi}\\
0&-\sin{\theta}e^{i\phi}&\cos{\theta}
\end{array}
\right),
\end{eqnarray}
and 
\begin{eqnarray}
U_{13}=\left(
\begin{array}{ccc}
\cos{\theta}&0&\sin{\theta}e^{-i\phi}\\
0&1&0\\
-\sin{\theta}e^{i\phi}&0&\cos{\theta}
\end{array}
\right).
\end{eqnarray}
Consequently, we have derived a novel neutrino mixing model, denoted as MTM1 and MTM2, which bears similarities to the well-known TM model. The TM model fails to simultaneously reproduce the best-fit values of the solar and reactor mixing angles. However, when employing $\alpha=-0.01483$ for MTM1 and $\alpha=-0.03708$ for MTM2, these models simultaneously predict the best-fit values for both angles. Moreover, the Dirac CP phase predicted by TM1 (TM2) mixing, represented as $\delta_{\rm{TM1}}$ ($\delta_{\rm{TM2}}$), closely aligns with the Dirac CP phase predicted by MTM1 (MTM2) $\delta_{\rm{MTM1}}$ ($\delta_{\rm{MTM2}}$), yielding nearly identical results.

Finally, we have shown that the neutrino mass matrix $M_{\rm{MTM1}}$ ($M_{\rm{MTM2}}$) that is obtained from $U_{\rm{MTM1}}$ ($U_{\rm{MTM2}}$) is invariant under $Z_2$ symmetry.
\vspace{1cm}

\end{document}